# Microwave attenuators for use with quantum devices below 100 mK

Jen-Hao Yeh,[1,2,a)] Jay LeFebvre,[1,2,b)] Shavindra Premaratne,[1,2] F. C. Wellstood,[2,3] and B. S. Palmer[1,2]

[1]*Laboratory for Physical Sciences, 8050 Greenmead Drive, College Park, Maryland 20740, USA*
[2]*Department of Physics, University of Maryland, College Park, Maryland 20742, USA*
[3]*Joint Quantum Institute, University of Maryland, College Park, Maryland 20742, USA*

To reduce the level of thermally generated electrical noise transmitted to superconducting quantum devices operating at 20 mK, we have developed thin-film microwave power attenuators operating from 1 to 10 GHz. The 20 and 30 dB attenuators are built on a quartz substrate and use 75 nm thick films of nichrome for dissipative components and 1 $\mu$m thick silver films as hot electron heat sinks. The noise temperature of the attenuators was quantified by connecting the output to a 3D cavity containing a transmon qubit and extracting the dephasing rate of the qubit as a function of temperature and dissipated power $P_d$ in the attenuator. The minimum noise temperature $T_n$ of the output from the 20 dB attenuator was $T_n \leq 53$ mK for no additional applied power and $T_n \approx 120$ mK when dissipating 30 nW. In the limit of large dissipated power ($P_d > 1$ nW), we find $T_n \propto P_d^{1/5.4}$, consistent with detailed thermal modeling of heat flow in the attenuators. *Published by AIP Publishing.*
[http://dx.doi.org/10.1063/1.4984894]

## I. INTRODUCTION

Since the seminal results by Nakamura *et al.*,[1] coherence times for superconducting qubits have risen by five orders of magnitude to values in excess of 100 $\mu$s.[2,3] These improvements are a result of new qubit designs, better techniques for operating devices and reading out the quantum state,[4–8] a greater understanding of energy relaxation mechanisms,[3,9–11] and better fabrication procedures.[3,12]

With these improvements, the devices have become much more sensitive to dephasing.[2,7,8,13,14] Dephasing is a result of fluctuations in the transition frequency of a quantum system, and common sources of dephasing include charge noise,[15] flux noise,[16] and critical current fluctuations.[17] However, in transmons with a single Josephson junction, dephasing due to flux noise is minimal, and the large ratio of the Josephson energy $E_J$ to the charging energy $E_C$ ensures that dephasing from low frequency charge noise is negligible.[6]

An additional source of dephasing for transmons occurs when the devices are coupled to a resonant microwave circuit or cavity that is used to measure the state of the qubit in the circuit QED architecture.[4,13,14,18–20] Due to coupling between the cavity and the qubit, a dispersive frequency shift of $2\chi$ is imparted to the qubit transition frequency for each photon stored in the cavity.[4,18,19,21] Random fluctuations in the number of photons in the cavity cause fluctuations in the qubit transition frequency, leading to dephasing.[13,14,20] In thermal equilibrium at temperature $T < 20$ mK, the average number of thermal photons $\bar{n}_{th}$ at 8 GHz would be $\bar{n}_{th} < 10^{-8}$, corresponding to an insignificant dephasing rate $\Gamma_\varphi$ for typical cavity-qubit coupling strength.

Instead, the coupling of the cavity to Johnson-Nyquist noise from the control and measurement microwave lines increases $\bar{n}_{th}$ orders of magnitude above $10^{-8}$ due to the lack of isolation from blackbody noise at temperatures larger than 20 mK. For thermalizing the signals, a series of attenuators or directional couplers are mounted at different temperature stages in the refrigerator. For example, we typically use a total attenuation of 70 dB in our superconducting qubit experiments, and ideally, this attenuation should reduce the noise to a minimum temperature of 45 mK at 8 GHz if the attenuators or directional couplers are well thermalized and have minimal self-heating.

In our experimental system,[22] we had found that noise on the cavity input line with an effective temperature as large as $T_n \approx 120$ mK drove the cavity so that the average number of photons was $\bar{n} \sim 0.1$.[22] The input line contained commercial microwave attenuators, not designed for millikelvin operation,[23] suggesting that the noise temperature was due to heating of the attenuator.

In this article, we improve upon the cooling of attenuators by designing, fabricating, and packaging the attenuators to operate at millikelvin temperatures. In Sec. II, we discuss the choice of materials, resistor configuration, and results from electrical and thermal simulations that went into the design of the attenuators. At temperatures below a few hundred millikelvin, dissipating electrical power in a thin film of normal metal can cause the electron temperature $T_e$ to increase substantially above the phonon temperature $T_p$.[24,25] This hot electron effect as well as the Kapitza phonon-phonon boundary resistances[26] were taken into account in the design and enabled us to decrease the effective photon noise temperature to $T_n \leq 53$ mK. In Sec. III, we present measurements of dephasing in a 3D transmon as a function of temperature and power $P_d$ dissipated in the attenuator, and we use this data to characterize the attenuator

[a)]Electronic mail: davidyeh@umd.edu
[b)]Current address: Department of Physics and Astronomy, University of California, Riverside, California 92521, USA.

performance. In Sec. IV, we conclude and make suggestions for further improvements.

## II. ATTENUATOR DESIGN AND FABRICATION

In addition to requiring good electrical and thermal properties, several other factors were taken into consideration for the design of our attenuators. Ideally, we wanted an attenuator that was relatively compact, could be fabricated using standard techniques, and was constructed from materials with physical properties that were well understood. Furthermore, to be compatible with many different types of superconducting quantum devices, we wanted simple, matched, dissipative attenuators with a flat microwave response between 1 and 10 GHz.

### A. Circuit layout

There were two main issues to be addressed in the circuit layout. The first had to do with the number of attenuator stages. For example, if one desires a total attenuation of 20 dB at 20 mK, should the attenuator be designed with one 20 dB attenuator, a series of two 10 dB attenuators, a series consisting of 10, 5, 2.5, and 2.5 dB attenuator sub-stages, or some other layout variation? Initial thermal simulations suggested little advantage between the different options, so for simplicity, we decided to develop a design with a series of 10 dB attenuators. To isolate each 10 dB stage from thermoelectric currents, we included an interdigitated capacitor between stages. The second main design choice was what type of attenuator circuit should be used. The two obvious candidates were T-pad and Π-pad designs.[27,28] In thermal simulations for a range of input powers, we found that the T-pad designs had a larger cooling power than the Π-pad designs, so we decided to proceed with a T-pad configuration embedded in a microwave coplanar waveguide geometry (CPW) (see Fig. 1).

The resistances $R_1$, $R_2$, $R_3$, and $R_4$ in our T-pad [see Fig. 1(a)] were chosen based on the desired amount of power attenuation $K^2$, impedance matching to the characteristic impedance $Z_0$ of the input and output microwave lines, and symmetry. With these constraints, the values of the resistors are

$$R_1 = R_2 = Z_0 \frac{K-1}{K+1} \quad (1)$$

and

$$R_3 = R_4 = Z_0 \frac{4K}{K^2 - 1}, \quad (2)$$

where $K = V_{\text{in}}/V_{\text{out}}$ is the ratio of the input voltage to the output voltage. For a 10 dB attenuator sub-stage ($K^2 = 10$) connected to a characteristic impedance of $Z_0 = 50\,\Omega$, one finds $R_1 = R_2 = 26.0\,\Omega$ and $R_3 = R_4 = 70.3\,\Omega$.

With the resistance values determined, the dimensions of each resistor were computed according to the sheet resistance of a resistive thin film. While physically larger resistors are thermally advantageous, the maximum size of the shunt resistors is limited by the microwave response. For a large resistor, its parasitic capacitance and inductance become important, and a lumped element model is not valid. Microwave simulations using ANSYS's high frequency software simulator (HFSS) confirmed that the response of the attenuator degraded at high frequencies when the physical size of the resistors got too large. To achieve a flat microwave response (less than 3 dB change) up to 12 GHz in the simulations, we decided upon a CPW geometry with a center conductor width $w_c = 0.8$ mm and a gap width $w_g = 0.26$ mm [see Fig. 1(b)]; parameters that obviously depend on the dielectric constant of the substrate (see Sec. II B).

### B. Material selection

We chose to use thin-film nichrome (NiCr) for the dissipative resistors in the attenuators. Nichrome (80% Ni and 20% Cr by weight) has a stable resistivity which only decreases by $\sim$10% when cooled to cryogenic temperature.[29,30] A thin film (75±5 nm) of nichrome was sputtered onto a single-crystal quartz substrate forming a layer with a sheet resistance of approximately $R_s = 27\,\Omega/\square$. Quartz was used because it has a small relative permittivity ($\epsilon_r = 3.9$), allowing the lumped element resistors in the attenuator to be physically larger without introducing too much parasitic capacitance. Quartz also has a reasonably high thermal conductivity (4.0 W m$^{-1}$ K$^{-1}$ at 1 K) for an insulating substrate at low temperatures.[31,32]

After patterning the NiCr elements via photolithography and wet etching, a 1 $\mu$m thick film of patterned Ag was deposited by electron beam evaporation. This Ag layer was used to connect the resistive elements to one another in the T-pad configuration, to conduct heat out of the dissipative nichrome regions, and to provide cooling of the electrons via an electron-phonon coupling.

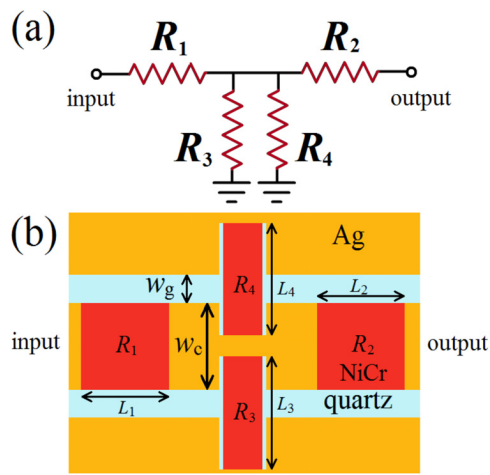

FIG. 1. (a) Lumped-element circuit model for a T-pad attenuator. (b) Top view of the layout for a single 10 dB attenuator cell. The 75 nm thick NiCr resistors are connected by a 1 $\mu$m thick silver film in a coplanar waveguide geometry (center conductor width $w_c = 0.8$ mm and gap width $w_g = 0.26$ mm) on a 0.5 mm thick quartz substrate.

### C. Thermalization model

Based on a total input microwave power $P_{\text{in}}$ to the attenuator, the dissipated power ($P_i$) in each resistor ($R_i$) was calculated using the lumped element model of the attenuator [see Fig. 1(a)]. For simplicity, the power $P_i$ was assumed to be uniformly dissipated throughout the volume of the $i$-th NiCr resistor. Figure 2 shows a simplified schematic of the heat flow in one of the resistor sections in the attenuator. Heat from $P_i$ is delivered to the NiCr electrons and flows to the NiCr phonons by electron-phonon coupling ($P_{e-ph}^{(NiCr)}$) or to the Ag electrons by electron conduction ($P_{e-e}^{(leads)}$). In the steady state

$$P_i = P_{e-ph}^{(NiCr)} + P_{e-e}^{(leads)}. \quad (3)$$

At millikelvin temperatures, the transfer of heat between the electrons and phonons in a thin metal film can be very poor. When electrical power is dissipated, it drives the electron temperature $T_e$ out of equilibrium with the phonon temperature $T_p$ and makes the electron temperature hot (i.e., $T_e \gg T_p$).[24,25] In our system, we model this electron effect using[25]

$$P_{e-ph} = V_m \Sigma \left( T_e^5 - T_p^5 \right), \quad (4)$$

where $P_{e-ph}$ is the net rate at which heat is transferred from the electrons to the phonons, $V_m$ is the volume of the metal, and $\Sigma$ is a material-dependent parameter for coupling between the electrons and phonons. For our design, we assumed $\Sigma_{Ag} = 10^8$ W m$^{-3}$ K$^{-5}$ for the Ag films[25] and $\Sigma_{NiCr} = 5 \times 10^8$ W m$^{-3}$ K$^{-5}$ for the NiCr resistors.

The transfer of heat from a thin film to a substrate can also be limited by phonon scattering at the interface. We model the effect of Kapitza boundary resistance using[26]

$$P_{ph-ph} = G_K(T) \Delta T, \quad (5)$$

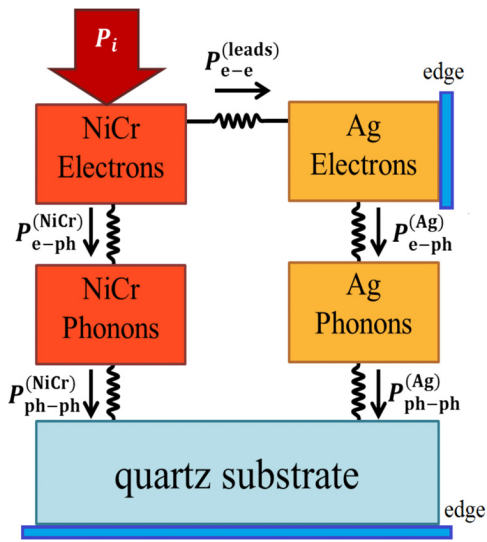

FIG. 2. Schematic of heat flow from electrical power $P_i$ dissipated in a NiCr resistor. This model, which includes hot electron effects ($P_{e-ph}$) and Kapitza boundary resistance effects ($P_{ph-ph}$), was used in a finite element model to simulate the electron and phonon temperatures of the NiCr and Ag films. The outer perimeter of the Ag film and the back of the quartz film (denoted edge in the figure) are assumed to be at the bath temperature $T_b$.

where $\Delta T$ is the difference in phonon temperatures between the two materials, $G_K(T) = A_K B T^3$ is the Kapitza boundary conductance, $A_K$ is the cross-sectional interface area, and we set $B_{Ag} = 1200$ W m$^{-2}$ K$^{-4}$ for the silver-quartz interface and $B_{NiCr} = 950$ W m$^{-2}$ K$^{-4}$ for the nichrome-quartz interface.[33,34] The phonon temperature at the bottom edge of the quartz substrate is assumed to be the same as the bath temperature $T_b$.

Heat can also flow out of the NiCr resistors into the Ag film by electron thermal conduction ($P_{e-e}^{(leads)}$). Once in the Ag layer, $P_{e-e}^{(leads)}$ flows to the Ag phonon bath [i.e., Eq. (4)] or diffuse laterally by thermal conduction. The outer perimeter of the Ag film, which is in contact with the copper attenuator package (see Fig. 2) using silver epoxy, is assumed to be at the bath temperature ($T_b$) of the plate on which it is mounted. The thermal resistance between NiCr and Ag is assumed to be negligible, and the conduction of heat in both the NiCr and Ag films is assumed to obey[35]

$$P_L = A_L \kappa_{th}(T) \nabla T, \quad (6)$$

where $P_L$ is the rate at which the heat flows via electron conduction through the cross sectional area $A_L$ of the film, $\kappa_{th}(T) = \kappa_e T$ is the temperature-dependent thermal conductivity, and $\nabla T$ is the temperature gradient. In our simulation, the coefficient $\kappa_e$ is set to 102 W m$^{-1}$ K$^{-2}$ for silver and 0.02 W m$^{-1}$ K$^{-2}$ for nichrome.[36]

To assess attenuator designs, simulations of the microwave response were performed using HFSS, and thermal simulations were performed using COMSOL Multiphysics. In the thermal simulations, the bath temperature $T_b$ was varied from 20 mK to 100 mK, and the input microwave power $P_{in}$ to the attenuator was adjusted from 0.1 nW to 100 $\mu$W. The geometry of the lines and resistors were varied to find the best design before fabricating the attenuators.

Figure 3(a) shows a finite element thermal simulation of the electron temperature $T_e$ of two 10 dB attenuators in series for our final design. In this simulation, the bath temperature $T_b$ is set to 20 mK, and the input power was set to 50 nW. Note that the simulation shows that the first NiCr resistor ($R_1$) reaches a nearly uniform electron temperature of 250 mK. This is much hotter than both the electrons in the connected Ag film and the NiCr phonon temperature, suggesting that both the thermal conductance of the NiCr layer acts as a bottleneck for heat leaving via electron conduction to the Ag layer, and the hot electron effects causes the electron temperature in the NiCr to be out of equilibrium with the phonon temperature. Further analysis of the simulation results shows that the electron temperature is only sensitive to two critical parameters $\Sigma_{NiCr}$ and $\kappa_{e,NiCr}$, and not sensitive to the other parameters ($\Sigma_{Ag}$, $\kappa_{e,Ag}$, $B_{Ag}$, and $B_{NiCr}$). While varying these other parameters by a factor of 10, the electron temperature only changes less than 2%. We note that the 50 nW input power is much larger than the instantaneous power used for the qubit control pulses (see Sec. III C).

From the spatially dependent temperatures found in the thermal simulations, we calculate the two-sided Johnson-Nyquist voltage noise power spectrum[37] from each resistor $R_i$ in the attenuator using the following integration

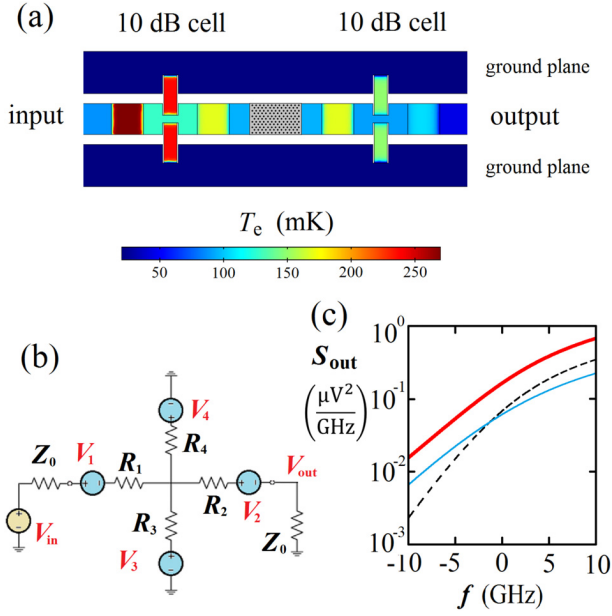

FIG. 3. (a) Finite-element simulation of the electron temperature $T_e$ of the NiCr and Ag films for a 20 dB attenuator with microwave input power $P_{in} = 50$ nW and a bath temperature $T_b = 20$ mK. The dotted gray region in the center trace between the two 10 dB cells denotes an interdigital capacitor, which was not included in the simulation to reduce simulation time. (b) Noise circuit model used with the simulated electron temperatures to calculate the output spectral density. (c) Two-sided voltage noise power spectral density at output calculated using the temperature distribution of resistors in (a), Eq. (7), and an input spectral density of noise with an effective noise temperature $T_{in} = 100$ mK. The thick red curve is the total output noise spectral density $S_{out}$ which yields $T_n = 124$ mK at 8 GHz by using Eq. (8). The black dashed curve is the contribution to $S_{out}$ from resistor $R_2$ in the second 10 dB cell, and the thin blue curve is the contribution to $S_{out}$ from resistors $R_3$ and $R_4$ in the second 10 dB cell.

$$S_i(f) = \frac{2hfR_s}{W_i} \int_0^{L_i} \frac{dx}{1 - \exp\left(-\frac{hf}{k_B T_e(x)}\right)}, \quad (7)$$

where $f$ is frequency, $R_s$ is the sheet resistance, $W_i$ is the width of the resistor, $L_i$ is the length of the resistor [see Fig. 1(b)], $T_e(x)$ is the simulated electron temperature distribution, and $x$ is the dimension parallel to the current flow along the resistor ($0 \leq x \leq L_i$). Note that Eq. (7) uses the fact that $T_e$ does not vary in the direction perpendicular to $x$.

With $S_i(f)$ determined for each resistive element in the attenuator circuit, the total output voltage noise power spectrum $S_{out}(f)$ from the attenuator is found using the circuit model in Fig. 3(b). In this model, each resistor $R_i$ has a voltage noise source $V_i$ in series, and the attenuator from the previous stage has noise spectrum $S_{in}(f)$ with the characteristic impedance $Z_0$. The thick red curve in Fig. 3(c) shows $S_{out}(f)$ from the simulation result shown in Fig. 3(a) with an input noise temperature of 100 mK to the 20 dB attenuator. While the resistor $R_1$ within each 10 dB cell dissipates the most power and has the largest temperature, the noise contribution from $R_1$ to the output of the cell is attenuated by the T-pad circuit. For this reason, the noise from $R_2$ [black dashed curve in Fig. 3(c)], and from both $R_3$ and $R_4$ [thin blue curve in Fig. 3(c)] contribute more to $S_{out}(f)$.

Note that since the resistors are at different temperatures when power is applied, the resulting output spectrum is, in general, non-thermal. Nevertheless, we define the effective output noise temperature of the attenuator at frequency $f$ as[37]

$$T_n(f) \equiv \frac{hf}{k_B} \left[\ln\left(\frac{S_{out}(+f)}{S_{out}(-f)}\right)\right]^{-1}, \quad (8)$$

which reduces to the frequency-independent temperature $T_n$ in thermal equilibrium. Using the finite element simulations for the temperature profile of each resistor, Eq. (7) for the noise power spectrum from each resistor and Eq. (8), we calculated the effective noise temperature output from specific attenuator designs.

Table I summarizes the parameters for our final design, which gave the lowest effective noise temperature at the attenuator output consistent with the materials, fabrication, and design constraints.

## III. ATTENUATOR PERFORMANCE WITH A SUPERCONDUCTING QUANTUM BIT

### A. Microwave response of attenuators

We fabricated 20 and 30 dB attenuators on 3-inch diameter single-crystal quartz wafers (stable temperature cut and double side polished)[38] that were later diced into individual 17.7 mm × 3.5 mm × 0.5 mm chips. The individual chips were packaged in an oxygen-free-high-conductivity Cu box (see Fig. 4). Silver epoxy[39] was used to provide electrical and thermal bonding between the box and the ground plane of the chip, and it was also used on the back of the chip to secure the chip to the box. Since the attenuators need to operate below the minimum recommended operating temperature of the epoxy (182 K),[39] failure of the epoxy bond was of concern. Using an inspection microscope to examine attenuators that underwent up to 10 thermal cycles did not reveal any cracks in the epoxy, and we also found no degradation in the microwave or thermal response. SubMiniature version A (SMA) connectors were bolted to the sides of the Cu box, and the SMA center pins were soldered with a Pb/Sn based solder to the center conductor of the attenuator chip for the input and output connections to the attenuator.

The microwave response of the attenuators was examined before assessing the thermal performance. Figure 5 shows the transmitted power through a 20 dB and a 30 dB attenuator measured at room temperature and at 5 K. At room temperature, the broadband attenuation in each attenuator was close to the designed value in the 1 to 10 GHz range, with a deviation of less than 3 dB. The transmission data at 5 K (solid black curves in Fig. 5) differs by about 3 dB from the 300 K data (dashed red curves), likely because of decreased attenuation from the stainless steel coax lines inside the Dewar which were only calibrated at room temperature, and therefore, not necessarily due to changes in the attenuator.

TABLE I. Design parameters of each 10 dB cryogenic attenuator cell in a coplanar waveguide (CPW) geometry. The resistances of the resistors are $R_1 = R_2 = 26.0\,\Omega$ and $R_3 = R_4 = 70.3\,\Omega$. The finger width and the gap width of the interdigital capacitor are 5 $\mu$m.

|  | Length (mm) | Width (mm) | Thickness ($\mu$m) | Material |
|---|---|---|---|---|
| Substrate | 17.740 | 3.500 | 500 | Quartz |
| Center pin of CPW | … | 0.800 | … | Ag |
| Gap of CPW | … | 0.260 | … | … |
| Ground pad | 17.740 | 1.090 | 1 | Ag |
| Resistors $R_1$ and $R_2$ | 0.831 | 0.800 | 0.075 | NiCr |
| Resistors $R_3$ and $R_4$ | 1.040 | 0.368 | 0.075 | NiCr |
| Interdigital capacitor | 0.800 | 0.800 | 1 | Ag |

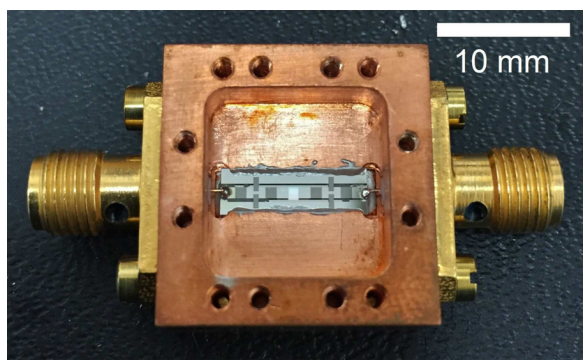

FIG. 4. Photograph of packaged 20 dB attenuator (copper lid not shown). Silver epoxy connects both sides of the ground plane of the attenuator chip to the copper package.

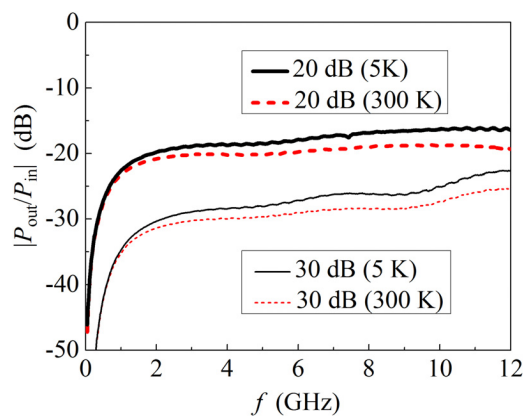

FIG. 5. The plot of the measured ratio of the magnitude of output transmitted power $P_{out}$ to input power $P_{in}$ versus frequency $f$ for a 20 dB attenuator (thicker curves) and a 30 dB attenuator (thinner curves) at 5 K (solid black curves) and at 300 K (dashed red curves). $P_{out}$ decreases below 1 GHz due to the power being reflected off the interdigital capacitor placed in the center trace between the 10 dB attenuator cells.

### B. Cryogenic system and the transmon qubit

By using a Leiden Cryogenics CF-450 dilution refrigerator, the effective noise temperature of the attenuators at millikelvin temperatures was assessed by measuring the dephasing rate of a 3D transmon qubit. Figure 6 shows a schema of the layout. A commercial off-the-shelf (COTS) 20 dB attenuator[40] was mounted on the 3 K plate on the input line. This was followed by our 30 dB attenuator, bolted to the cold plate (75 mK) of the refrigerator. The input line next went to a dc block followed by our 20 dB attenuator, and then a 12 GHz low pass filter[41] before going to the input port

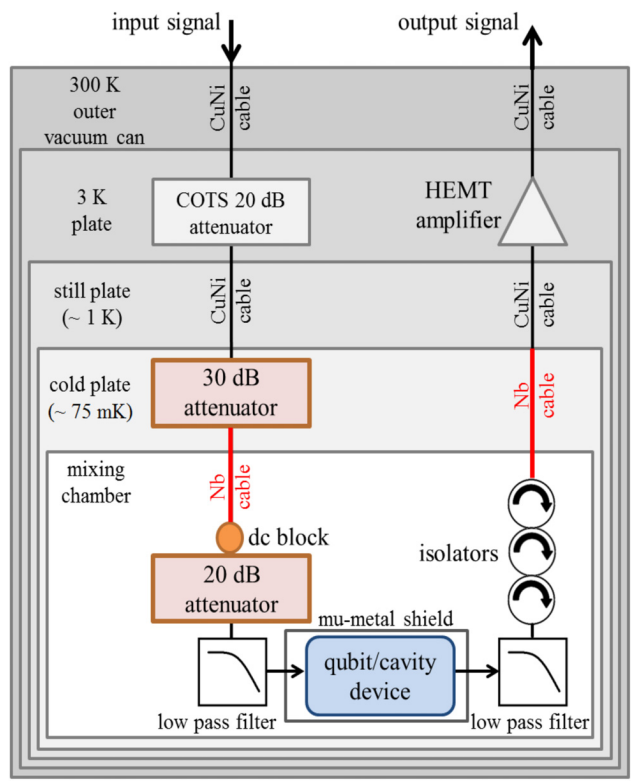

FIG. 6. Schema showing the placement of components for filtering and thermalizing the microwave signals in the dilution refrigerator.

of the cavity, all of which were mounted on the mixing chamber.

The output port of the cavity was connected to a 12 GHz low pass filter[41] and three microwave isolators[42] mounted on the mixing chamber (see Fig. 6). After the final isolator, the signal was amplified at 3 K with a low-noise high-electron-mobility-transistor (HEMT) amplifier[43] before going to room temperature. The microwave coax cables connecting the components between different temperature stages were CuNi or Nb cables (see Fig. 6).[44] Two cryogenic mu-metal cylinders[45] surrounded the cavity. Radiation cans, standard with the CF-450, were mounted on each stage of the refrigerator. We note that neither IR filters nor IR absorbing coatings of the shields[46] were used in this system, and without these precautions, we have observed coherence times up to 100 $\mu$s in previous 3D transmon devices.[47]

The three-dimensional superconducting aluminum cavity had a TE$_{101}$ fundamental mode at $f_c = \omega_c/2\pi = 7.924$ GHz and contained a fixed frequency transmon qubit.[7] For exciting and measuring the qubit, the input and output lines were connected to two SMA launchers in the side of the cavity. We note that the effective temperature $T_r$ of the cavity photons is determined by the noise on the cavity input and output ports, weighted by the coupling factors $1/Q_{in}$ and $1/Q_{out}$, respectively. Since the goal was to measure the effective noise temperature $T_n$ of the signals from the last attenuator on the input line, we designed the coupling of the cavity to the input port to be much larger than the coupling to the output port. From the reflectance and transmission measurements at room temperature, the following parameters were determined $Q_{in} = 3700 \ll Q_{out} \cong 10^5$. In this limit, the noise from the input line dominates and $T_r \cong T_n$.

The qubit was fabricated on a sapphire substrate and consisted of a single Al/AlO$_x$/Al Josephson junction shunted by two large Al pads, which reduced its charging energy to $E_C/h = 190$ MHz and provided coupling to the TE$_{101}$ fundamental mode of the cavity. The undriven qubit-cavity system can be described by the dispersive Jaynes-Cummings Hamiltonian[6,22,48]

$$\widehat{H}_{JC} = \hbar\omega_c \widehat{N} + \frac{\hbar}{2}\left(\omega_q + 2\chi\widehat{N}\right)\widehat{\sigma}_z, \quad (9)$$

where $\widehat{\sigma}_z$ is the Pauli matrix operating on the state of the qubit, $\widehat{N}$ is the cavity photon number operator, $\omega_q/2\pi = 6.550$ GHz is the qubit transition frequency, and $\chi/2\pi = -5.0$ MHz is the measured dispersive shift resulting from cavity-qubit coupling.

### C. Qubit dephasing and noise temperature

The transverse coupling between the cavity and qubit produces a shift of $2\chi$ in the qubit transition frequency for each additional photon stored in the cavity. Consequently, the qubit dephases if the photon number fluctuates.[14,18–20] We probe our qubit-cavity system at the $N = 0$ qubit frequency, in the limit of small photon number ($n \ll 1$), weak damping (i.e., $\kappa \ll \omega_c$ where $\kappa$ is the cavity decay rate), weak coupling (i.e., $\chi \ll \omega_c$), and strong dispersion (i.e., $|\chi| > \kappa > \Gamma_1$, where $\Gamma_1/2\pi \cong 0.06$ MHz is the qubit relaxation rate).[21] In this limit, if the dephasing of the qubit is dominated by the thermal photons with an average value of $\bar{n}_{th}$ and other dephasing sources can be neglected, the dephasing rate for the qubit reduces to simply[13,20]

$$\Gamma_\varphi = \kappa \bar{n}_{th}. \quad (10)$$

Here, the cavity decay rate, $\kappa/2\pi = (1.4 \pm 0.1)$ MHz, was extracted by fitting to both quadratures of the transmitted resonance of the dressed cavity at low temperatures. We note that the resonance had a quality factor of $Q_L = \frac{\omega_c}{\kappa} = 5700$ which differs from the room temperature value of $Q_{in} = 3700$, an effect that we attribute to the thermal contraction of the SMA pin. In our analysis, we have neglected dephasing due to TE$_{103}$ mode[13] at 13.2 GHz since this mode's frequency was higher than the 12 GHz cutoff frequency of the low pass filters and the qubit-cavity coupling was much weaker ($\chi_{103}/2\pi = -0.6$ MHz).

From the measurements of the qubit relaxation time $T_1$ and spin-echo coherence time $T_2$, the qubit dephasing rate using $\Gamma_\varphi = T_2^{-1} - (2T_1)^{-1}$ was extracted. The cavity readout pulses were 4 $\mu$s long and repeated every 400 $\mu$s. For placing the qubit in its excited state for the relaxation measurement, we used a $\pi$-pulse with a 100 ns duration. Figure 7(a) shows the measured $T_1$ (black squares) and $T_2$ (red circles) at different mixing chamber temperatures $T_m$. For $T_m < 130$ mK, $T_1$ was roughly constant; the scatter in $T_1$ below 100 mK was a result of two separate cooldowns. Above 130 mK, the thermal activation of quasiparticles results in a rapid decrease in $T_1$ as $T_m$ increases. On the other hand, the coherence time $T_2$ begins to noticeably decrease above 50 mK due to the photons populating the cavity.

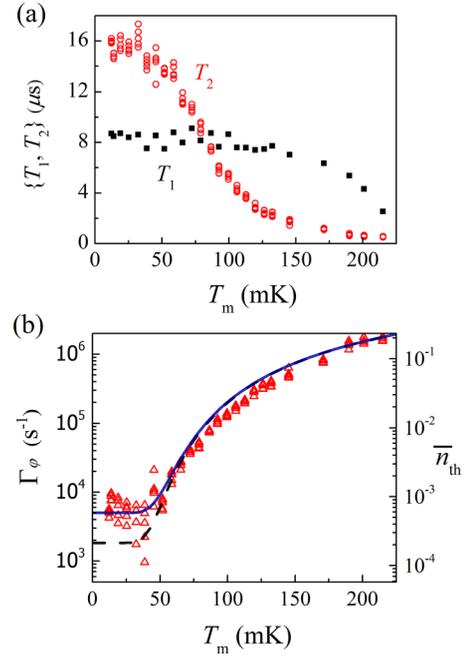

FIG. 7. (a) Measured relaxation time $T_1$ (black squares) and coherence time $T_2$ (red circles) of the qubit versus the mixing chamber temperature $T_m$. (b) Extracted dephasing rate $\Gamma_\varphi$ (red triangles) and the corresponding average number of thermal photons $\bar{n}_{th}$ by Eq. (10). The two lines are the expected dephasing rate when the cavity photons are in equilibrium with $T_m$, plus an offset dephasing rate $\Gamma_\varphi^{min} = 1.8 \times 10^3$ s$^{-1}$ (dashed black), which is the expected lower bound in the noise temperature of $T_n = 45$ mK due to noise from higher stages, and an average measured value $\Gamma_\varphi^{min} = 5 \times 10^3$ below 40 mK (solid blue).

The extracted dephasing rate $\Gamma_\varphi$ (red triangles) as a function of $T_m$ is shown in Fig. 7(b). For comparison, the two curves in Fig. 7(b) are calculated using

$$\Gamma_\varphi = \kappa\left[\exp\left(\frac{hf_c}{k_B T_r}\right) - 1\right]^{-1} + \Gamma_\varphi^{min}, \quad (11)$$

where the first term is Eq. (10), assuming that the effective temperature of the cavity $T_r$ is in equilibrium with the mixing chamber temperature ($T_r = T_m$) and $\bar{n}_{th}$ obeys Bose-Einstein statistics, and $\Gamma_\varphi^{min}$ is a constant offset in the dephasing describing, where the system deviates from a simple thermodynamic model. The dashed black curve uses $\Gamma_\varphi^{min} = 1.8 \times 10^3$ s$^{-1}$ corresponding to an expected lower bound in noise temperature of $T_n = 45$ mK which is limited by noise from the higher temperature stages and the amount of attenuation in the system. This limit would be reached if each attenuator properly attenuated the noise from the previous stage, and the temperatures of its resistors were at the bath temperature of the stage. The solid blue curve uses the average measured dephasing rate $\Gamma_\varphi^{min} = (5 \pm 2) \times 10^3$ below $T_m < 40$ mK for its offset. Using Eq. (10), this average rate of dephasing corresponds to $\bar{n}_{th} < 8 \times 10^{-4}$ photons and $T_r < 53$ mK which is close to the lower bound and a factor of two smaller in temperature than what we obtained previously using COTS attenuators at millikelvin temperatures.[22]

The effective temperature of noise from the output line was also independently measured by making ports with $Q_{in} = 3.3 \times 10^5 \gg Q_{out} = 2200$. On a device similar to the one presented here, the dephasing rate measurements placed an upper bound in the effective temperature of the isolator in the output line to be $T_{iso} < 40$ mK.

We also measured the dephasing when the temperature of the attenuator was increased above $T_m$ by applying a steady tone at $f = 1$ GHz (far detuned from the qubit and cavity frequencies) to the input line. Figure 8(a) shows the measured $T_1$ and $T_2$ versus power $P_d$ dissipated in the 20 dB attenuator (on the mixing chamber) with the temperature of the mixing chamber at $T_m = 14$ mK (blue triangles) at low power. We note that at the maximum dissipated power $P_d = 250$ nW, the mixing chamber temperature increased from 14 mK to 40 mK and from 72 mK to 83 mK, respectively. The dissipated power in the 30 dB attenuator bolted to the cold plate (see Fig. 6) was 1000 times larger than the dissipated power $P_d$ in the 20 dB attenuator, causing the cold plate temperature to increase from 75 mK to 198 mK. For comparison, we also show results with heater power applied to the mixing chamber so that $T_m$ was 72 mK (red circles) at low power. In this situation, $P_d = 250$ nW caused $T_m$ to increase to 83 mK due to the cold plate temperature increasing from 87 mK to 212 mK.

We also note that in addition to the steady heating tone, the qubit probing tone dissipated 10 pW during the 100 ns qubit $\pi$-pulse in the final 20 dB attenuator, and the cavity readout tone dissipated 0.5 nW during the 4 $\mu$s readout pulse. Figure 8(b) shows the effective noise temperature $T_n$ calculated from the dephasing rate $\Gamma_\varphi$ extracted from $T_1$ and $T_2$. Above approximately $P_d = 1$ nW, the noise temperature begins to increase with a power dependence $T_n \propto P_d^{1/5.4}$, a power law which suggests that the cooling of the electrons in the attenuator is dominated by the electron-phonon coupling.

Finally, we can compare our measured $T_n$ versus $P_d$ to results from the finite element simulations and the model described in Sec. II C. In our model, we took into consideration the thermalization of the attenuators on the 3 K and cold plates as well, so that the input noise of the final 20 dB attenuator included contributions from the output noise from the other attenuators in the line. We found good agreement with the $T_n$ versus $P_d$ data [see Fig. 8(b)] by setting the electron-phonon coupling strength to $\Sigma_{NiCr} = 5 \times 10^8$ W m$^{-3}$ K$^{-5}$ for the NiCr film. This value for $\Sigma_{NiCr}$ compares well with previously measured results on different materials.[25,49] As Fig. 8(b) shows, the main difference between the fit and data is in the saturation of $T_n$ below 3 nW for the $T_m = 14$ mK data. The saturation of $T_n$ at 53 mK could be due to a small systematic error in our determination of $\Gamma_\varphi$, another source of dephasing, an unknown source of thermal power heating the attenuator, or the attenuator chip being warmer than $T_m$.

## IV. CONCLUSION

In summary, we have developed broadband cryogenic thin-film microwave attenuators with an effective noise temperature below 100 mK when the dissipated power was less than 10 nW. In comparison with the worst case observations with commercial attenuators, we observed a factor of two decrease in the effective temperature to $T_n < 53$ mK and two orders of magnitude decrease in the average number of thermal photons to $\bar{n}_{th} < 8 \times 10^{-4}$ photons. Measurements of the population of the second excited state of the qubit also showed that its effective temperature was less than 50 mK. We note that for these measurements, the qubit-cavity system was specifically designed with $Q_{in} \ll Q_{out}$, so that the system was sensitive to the noise temperature of the attenuator. For typical qubit experiments, one arranges the coupling so that $Q_{in} \gg Q_{out}$, and this leads to less cavity photon noise from the input line. For example, if $Q_{in} = 3.3 \times 10^5 \gg Q_{out} = 2200$, then the cavity photon contributed from the thermal noise of the input line becomes $5 \times 10^{-6}$.

Our attenuator design process took into account the microwave transmission characteristics as well as hot electron and Kapitza boundary effects. In finite-element thermal simulations, we found that the dominant heating effect was due to the hot electrons and by matching the simulations to the measurements, we extracted an electron-phonon coupling factor $\Sigma_{NiCr} = 5 \times 10^8$ W m$^{-3}$ K$^{-5}$ for nichrome. Finally, our results suggest that further improvements in the cooling power of the attenuator could be achieved by significantly increasing the effective volume of the nichrome so that the effective coupling between the electrons and phonons in the attenuator is increased.

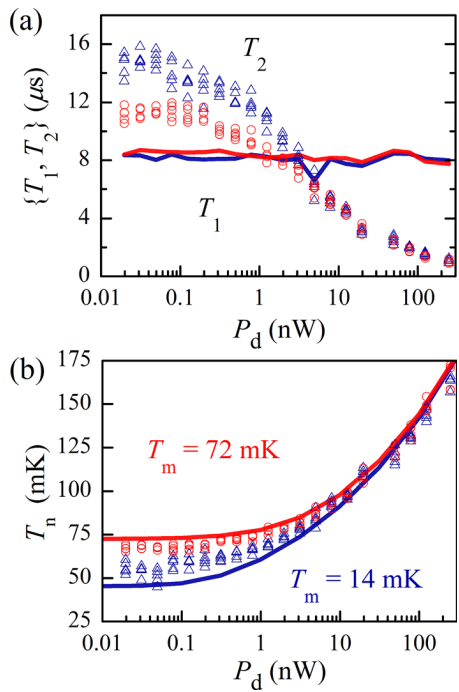

FIG. 8. (a) Measured $T_1$ (blue and red solid curves) and $T_2$ (blue triangles and red circles) versus dissipated power $P_d$ through the 20 dB attenuator on the mixing chamber when the temperature of the mixing chamber was 14 mK (blue) and 72 mK (red), respectively. (b) Effective noise temperature $T_n$ versus dissipated power $P_d$ in the 20 dB attenuator when the temperature of the mixing chamber was 14 mK (blue triangles) and 72 mK (red circles). The solid curves are calculated from the finite-element simulations described in Sec. II C and using an electron-phonon coupling constant $\Sigma_{NiCr} = 5 \times 10^8$ W m$^{-3}$ K$^{-5}$ for the NiCr film.

## ACKNOWLEDGMENTS

F.C.W. acknowledges support from the Joint Quantum Institute and the State of Maryland through the Center for Nanophysics and Advanced Materials.